%% file: main.tex
\def\year{2020}\relax
\documentclass[letterpaper]{article} 
\usepackage{aaai20}  
\usepackage{times}  
\usepackage{helvet} 
\usepackage{courier}  
\usepackage[hyphens]{url}  
\usepackage{graphicx} 
\usepackage{graphicx}  
\usepackage{booktabs}
\usepackage{amsmath}
\usepackage{amssymb}
\usepackage{multicol}
\begin{document}

\include{abstract}
\include{supplementary}

\end{document}

%% file: abstract.tex
\title{Algorithmic Bias in Recidivism Prediction: A Causal Perspective}
\author{Aria Khademi\textsuperscript{\rm 1} and Vasant Honavar\textsuperscript{\rm 1}\\ 
\textsuperscript{\rm 1}College of Information Sciences and Technology\\ 
The Pennsylvania State University\\
khademi@psu.edu 
}

\maketitle

\begin{abstract}
ProPublica's analysis of recidivism predictions produced by Correctional Offender Management Profiling for Alternative Sanctions (COMPAS) software tool for the task, has shown that the predictions were racially biased against African American defendants. We analyze the COMPAS data using a causal reformulation of the underlying algorithmic fairness problem. Specifically, we assess whether COMPAS exhibits racial bias against African American defendants using FACT, a recently introduced causality grounded measure of algorithmic fairness. We use the Neyman-Rubin potential outcomes framework for causal inference from observational data to estimate FACT from COMPAS data. Our analysis offers strong evidence that COMPAS  exhibits racial bias against African American defendants. We further show that the FACT estimates from COMPAS data are robust in the presence of unmeasured confounding.
\end{abstract}

\section{Introduction}
\noindent There is growing concern that AI technologies can perpetuate or amplify undesirable bias or discrimination based on race, gender, and other protected social attributes. An example is the COMPAS software used by the United States Judiciary to predict the likelihood of recidvism for defendants based on their characteristics and past criminal record. ProPublica's analysis of the COMPAS tool \cite{angwin2016propublica} spurred extensive debate on whether the software was biased against African American defendants.

There have been many attempts to formalize various notions of algorithmic fairness \cite{barocas-hardt-narayanan}. Of particular interest are notions of fairness that require that individuals do {\em not} experience differences in outcomes (e.g., recidivism score) {\em caused by} factors that are outside their control (e.g., race). Recent work has shown that tests of fairness expressed solely using the joint distribution \cite{hardt2016equality} of the observed variables are  incapable of detecting unfairness. Hence, there is a growing interest in algorithmic fairness criteria that {\em causally} link protected attributes with the outputs (e.g., decisions, predictions) of the algorithm \cite{barocas-hardt-narayanan,khademi2019fairness}. The key intuition behind such fairness criteria is that the question ``Is the decision discriminatory with respect to a protected attribute?'' can be reframed as: ``Does the protected attribute have a causal effect on the decision?'' Answering such a question is complicated by the fact that these factors can be meaningfully related to other characteristics that may be relevant in determining what is fair, and requires careful application of state-of-the-art tools for estimating causal effects from observational data.

We assess whether COMPAS exhibits racial bias against African American defendants using FACT, a recently introduced explicitly causal measure of algorithmic fairness \cite{khademi2019fairness}, using the Neyman-Rubin potential outcomes framework  \cite{rubin2005causal}. Our analysis offers robust evidence that COMPAS exhibits racial bias against African American defendants. 

\begin{table*}[t]
	\caption{Results of matching on the COMPAS data. Estimate of FACT is denoted by $\hat{\gamma}$. Statistical significance level is $\alpha=0.05$.}
	\centering
	\begin{tabular}{lcccccc}
		\toprule
		COMPAS dataset &  &   &  &  &   &  \\
		\midrule
		Matching method &   \# of Treated Matches  &  \# of Control Matches   &    \(\overline{D}^{m}_{a, a^\prime}\)   &   \(\hat{\gamma}\)  &   Standard Error  &  P-value  \\
		\midrule
		NNM               & $ 1893  $     & $ 780 $    & $ 0.0002  $   & $ 0.734  $     & $ 0.258 $    & $ 0.004 $ \\
		NNMPC               & $ 1893  $     & $910 $    & $ 0.0123 $   & $ 0.251  $     & $ 0.222 $    & $ 0.257 $ \\
		MMMPC               & $ 1893  $     & $852 $    & $ 0.0073 $   & $ 0.331 $     & $ 0.292 $    & $ 0.253 $ \\
		FM               & $ 1893  $     & $ 1447 $    & $ 0.0002  $   & $ 0.624  $     & $ 0.223 $    & $ 0.005 $  \\
		\bottomrule   
	\end{tabular}
	\label{table:fact}
\end{table*}

\section{Methods} \label{Methods}
Denote each individual $i$ with $(\tilde{X}_i, A_i, Y_i)$ where $\tilde{X}$ is the vector of non-protected attributes, $A \in \{a, a^\prime\}$ is race, and $Y$ is the likelihood that COMPAS would predict recidivism ($Y=1$) or non-recidivism ($Y=0$). Let $Y_i^{(a)}$ be the {\em potential outcome} of individual $i$, if they had race $a$. For each individual, either $Y_i^{(a)}$ or $Y_i^{(a^\prime)}$ is observable. We use a causal notion of fairness, namely, fair in average causal effect on the treated (FACT) \cite{khademi2019fairness}: A decision function $h: \mathcal{X} \times \mathcal{A} \to \mathcal{Y}$ is fair on average over individuals sharing a certain race if $\mathbb{E}[Y_i^{(a)} - Y_i^{(a^\prime)} \, | \, A_i = a] = 0$.

We estimate FACT using the state-of-the-art matching based methods for causal inference  \cite{stuart2010matching}, i.e., for each African American defendant (we observe $Y_i^{(a)}$), we find their most similar ``match'' in terms of non-protected attributes among White defendants (and hence estimate $Y_i^{(a^\prime)}$). We use the following matching methods within the R package MatchIt (version 3.0.2) \cite{ho2011matchit}: {\bf (i)} Nearest neighbor matching (NNM), {\bf (ii)} Nearest neighbor matching with propensity caliper (NNMPC), {\bf (iii)} Mahalanobis metric matching with propensity caliper (MMMPC), and {\bf (iv)} Full matching (FM), all according to the parameters specified in \cite{khademi2019fairness}.

To measure goodness-of-matches, we examined {\bf (i)} absolute value of standardized difference in means of the treated (race $a$) and controlled (race $a^\prime$) in terms of the distance measure (propensity score), before ($\overline{D}_{a, a^\prime}$) and after ($\overline{D}^{m}_{a, a^\prime}$) matching, and {\bf (ii)} jitter plots and histograms of the distribution of propensity scores after matching. For high quality matches, $\overline{D}^{m}_{a, a^\prime}$ must be close to 0. As a result of the matching process, each individual is assigned a weight. Subsequently, we run the weighted regression $\mathbb{E}[Y^{(A)}] = \delta + \gamma A + \tilde{\theta}^\top \tilde{X}$ {\em on the matched data set} (having dropped the data points for which no match is found) and obtain $\hat{\gamma}$ as the estimated causal effect of $A$ on $Y$ measured by FACT.

In the absence of unmeasured confounding, estimates of FACT are doubly robust if either the matching model or the subsequent regression model are correct \cite{ho2011matchit}. To test for the effect of unmeasured confounding on our estimates of FACT, we run sensitivity analysis (SA) with the R package {\em rbounds} (version 2.1) \cite{keele2010overview}. We expose our estimates to a $\Gamma$ factor of unmeasured confounding and measure the change in significance of estimates (see \cite{khademi2019fairness,rosenbaum2005sensitivity} and Supplementary S1 for details).

\section{Experiments}

\subsection{Data}
The COMPAS data offer 2 years of data (2013-2014) from the COMPAS software tool. The question is whether COMPAS predicts different rates of recidivism for African Americans compared to Whites (all other things being equal). We designated African Americans as treated ($A=1$) and Whites as control ($A=0$). The binary outcome $Y$ is  the COMPAS  prediction ($Y=1$ indicating recidivism). We used the ``Violent'' data pre-processed using the procedure used by ProPublica yielding 3373 data points.\footnote{\url{https://github.com/propublica/compas-analysis}}

\subsection{Fairness Analysis Using FACT}
We estimated the causal effect of race on COMPAS outcome using the techniques described in Section Methods. FM yielded the highest number of matched data points with the lowest $\overline{D}^{m}_{a, a^\prime}$ (see Table \ref{table:fact}) and hence highest quality of matches (see Supplementary S2 for details).

The FACT estimates are summarized in  Table \ref{table:fact}. We were able to reject the null hypothesis $H_0: \gamma = 0$ (in the case of NNM and FM) which suggests that the recidivism scores predicted by COMPAS exhibit racial bias against African Americans. We speculate that the propensity caliper in NNMPC and MMMPC disregards some data points that are important in rejecting $H_0$. In the case of FM, odds of the COMPAS software predicting that African American defendants would recidivate after release is $\exp(0.624) \approx 1.87$ times that of White defendants. This result is in agreement with previous work, e.g., \cite{angwin2016propublica}. 

\subsection{Robustness to Unmeasured Confounders} We ran SA with $\Gamma$ ranging from $1$ to $10$. The larger $\Gamma$, the bigger the exposure to unmeasured confounders. Our estimates of NNM, NNMPC, MMMPC, and FM were robust to unmeasured confounding up to $\Gamma$s of $9$, $7.5$, $8$, and $5.5$, respectively. We conclude that our FACT estimates are robust to unmeasured confounders.

\bibliographystyle{aaai}
\bibliography{references.bib}

%% file: supplementary.tex


\section{S1: Impact of Unmeasured Confounding}
We are interested in measuring the causal effect of a sensitive attribute $A \in \{a, a^\prime\}$ (e.g., race) on an outcome $Y$ (e.g., recidivism score). For that, we must contrast $Y_i^{(a)}$ (i.e., outcome of the treated) and $Y_i^{(a^\prime)}$ (i.e., outcome of the controlled). However, the fundamental problem of causal inference states that either $Y_i^{(a)}$ or $Y_i^{(a^\prime)}$ are observable, but {\em not both} \cite{holland1986statistics}. The purpose of matching is to observe either, e.g., $Y_i^{(a)}$ and estimate the other, e.g., $Y_i^{(a^\prime)}$. The estimation is done by a matching model that finds the ``closest'' (in terms of some distance measure with respect to their non-protected attribute) individual $j$ to person $i$, and takes $Y_j^{(a^\prime)}$ as an {\em estimate} for $Y_i^{(a)}$ (see \cite{stuart2010matching} for a review on different matching models). Estimates obtained using matching are unbiased (if the matching model is correct) in the absence of unmeasured confounding \cite{ho2011matchit}. To test for the effect of unmeasured confounding on the obtained estimates, one must run sensitivity analysis (SA).

Suppose data points \(i\) and \(j\) are matched using a matching method. Suppose \(\Gamma\) is the odds ratio of \(i\) and \(j\) receiving a treatment. If matching is \emph{perfect} and there is no hidden bias, then \(\Gamma = 1 \), resembling a randomized controlled trial. If, however, matching is impacted by a hidden bias introduced by unmeasured confounders, then \(\Gamma > 1\) (or \(\Gamma < 1\)) after matching, indicating that the data point \(i\) is more (or less) likely to receive treatment as compared to the matched data point \(j\). This is a consequence of not having controlled for the hidden bias in the matching process. We perform SA \cite{rosenbaum2005sensitivity,liu2013introduction,jung2018algorithmic} to investigate the degree to which the unmeasured confounders impact \(\hat{\gamma}\). 

Let \(O_i\) and \(O_j\) be the odds of receiving the treatment for data points \(i\) and \(j\), respectively. Then, we have:

\begin{equation}
	\frac{1}{\Gamma} \leq \frac{O_i}{O_j} \leq \Gamma.
\end{equation}

SA proceeds by first assuming that \(\Gamma = 1\) (i.e., no hidden bias). Then, it increases the value of \(\Gamma\) (e.g., \(1, \ldots, 5\)), thus mimicking the presence of hidden bias, and examines the resulting changes to statistical significance of \(\hat{\gamma}\). Analysis using Wilcoxon's signed rank test for continuous outcomes and McNemar's test for binary outcomes provides lower and upper bounds for the p-value of \(\hat{\gamma}\). The \(\Gamma\) at which the significance of the upper bound for the p-value would change (e.g., from \(< 0.05\) to \(> 0.05\)) is the point at which \(\hat{\gamma}\) is no longer robust to hidden bias. We ran SA using the R package \emph{rbounds} (version 2.1) \cite{keele2010overview}.

\begin{figure}[t]
    \centering
    \includegraphics[width=\linewidth]{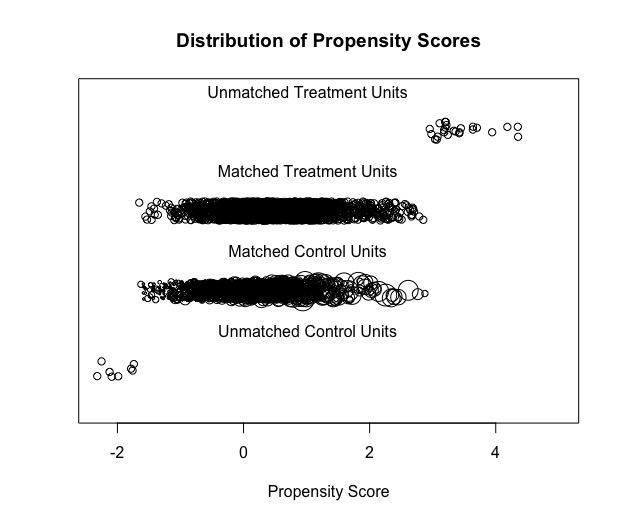}
    \caption{Jitter plot of the distribution of propensity scores (on the linear logit scale) of data points after FM. Each circle is a data point with its area being proportionate to the weight assigned to it. Blacks are treated and Whites are control.}
    \label{fig:jitter}
\end{figure}

\section{S2: Examining Goodness-of-Match}
Matching methods are only reliable if high quality matches are obtained. Following \cite{rubin2001using,stuart2010matching}, we assure high quality matches by examining the following:

\begin{enumerate}
    \item Absolute value of standardized difference in means of the treated and controlled in terms of the distance measure, before ($\overline{D}_{a, a^\prime}$) and after ($\overline{D}^{m}_{a, a^\prime}$) matching. To achieve high quality matches, $\overline{D}^{m}_{a, a^\prime}$ must be ideally zero.
    \item Jitter plots and histograms of the distribution of propensity scores before and after matching. The distributions of treated and controlled must be similar to each other after matching.
\end{enumerate}

\begin{figure}[!h]
    \centering
    \includegraphics[width=\linewidth]{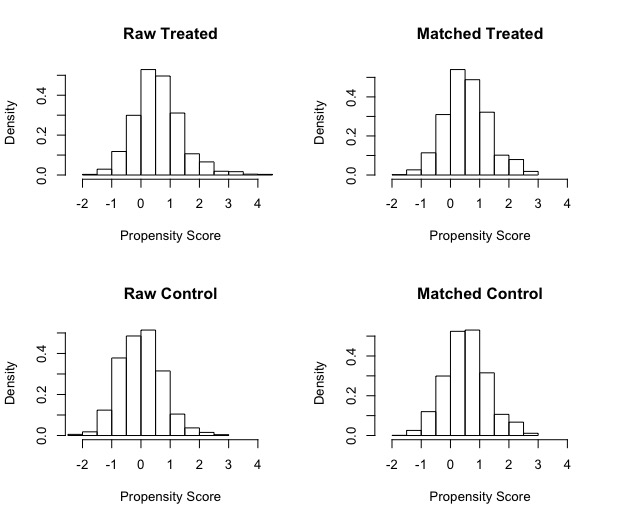}
    \caption{Histogram of the distribution of propensity scores (on the linear logit scale) of the data points after matching. Blacks are treated and Whites are control.}
    \label{fig:hist}
\end{figure}

We measured the quality of matches after matching and observed that $\overline{D}_{a, a^\prime}=0.5776$ while $\overline{D}^{m}_{a, a^\prime}$ is close to zero (see Table \ref{table:fact}). We further present results of our analyses on the quality of matches after FM in Figures \ref{fig:jitter} and \ref{fig:hist}. 

\newpage
In both figures, African American defendants are viewed as being treated and White defendants are viewed as being controlled. We observe that the distribution of the propensity scores of treated and controlled are much more similar to each other after matching than those before matching. Hence, we concluded desirable quality of matches has been achieved. After ensuring that the matches are of  high quality, and assigning appropriate weights to the data points (see \cite{stuart2010matching} for details of weighting), we proceeded to estimate FACT.